\documentclass[12pt]{article}

\usepackage{amsmath}
\usepackage{amssymb}
\usepackage{graphicx}
\usepackage{color}

\newcommand{\eps}{\epsilon}
\newcommand{\del}{\partial}
\newcommand{\lam}{\lambda}

\newcommand{\epsb}{\bar{\epsilon}}

\newcommand{\phib}{\bar{\phi}}
\newcommand{\fb}{\bar{F}}
\newcommand{\zb}{\bar{z}}
\newcommand{\delb}{\bar{\partial}}

\newcommand{\delp}{\partial_+}
\newcommand{\delm}{\partial_-}
\newcommand{\epsp}{\eps_+}
\newcommand{\epsm}{\eps_-}
\newcommand{\epsbp}{\epsb_+}
\newcommand{\epsbm}{\epsb_-}
\newcommand{\lamp}{\lam^+}
\newcommand{\lamm}{\lam^-}
\newcommand{\chip}{\chi^+}
\newcommand{\chim}{\chi^-}

\newcommand{\nn}{\nonumber}

\newcommand{\beq}{\begin{eqnarray}}
\newcommand{\eeq}{\end{eqnarray}}
\newcommand{\non}{\nonumber\\}
\newcommand{\nom}{\nonumber}

\newcommand{\ep}{\varepsilon}

\newcommand{\so}{&&}

\newcommand{\ba}{\begin{eqnarray}}
\newcommand{\ea}{\end{eqnarray}}
\newcommand{\bep}{\bar{\epsilon}}
\newcommand{\ps}{\psi}
\newcommand{\bps}{\bar{\psi}}
\renewcommand{\ep}{\epsilon}
\newcommand{\el}{\ell}
\newcommand{\de}{\delta}

\begin{document}


\thispagestyle{empty}

\begin{flushright}
 KUNS 1912 \\
\end{flushright}
\vspace{20mm}
\begin{center} 
{\bf \Large $N=2$ Supersymmetric Sigma Models and D-branes}
\vspace{20mm}

  {\bf Noriko Nakayama}
\footnote{\it  e-mail address: nakayama@phys.h.kyoto-u.ac.jp} \\
\vspace{8pt} 
{\it Graduate School of Human and Environmental Studies, Kyoto University, Kyoto 606-8501, Japan}\\ 
\vspace{10pt} 
and \\ 
\vspace{10pt} 
  {\bf Katsuyuki Sugiyama}
\footnote{\it  e-mail address: sugiyama@phys.h.kyoto-u.ac.jp} \\ 
\vspace{8pt}
{\it Graduate School of Science, Kyoto University, Kyoto 606-8501, Japan} 

\vspace{40pt}

{\bf Abstract}\\[5mm]
{\parbox{13cm}{\hspace{5mm}

We study D-branes of $N=2$ supersymmetric sigma models. 
Supersymmetric nonlinear sigma models with $2$-dimensional target space have D0,D1,D2-branes, 
which are realized as A-,B-type supersymmetric boundary conditions on the worldsheet. 
When we embed the models in the string theory, 
the K\"ahler potential is restricted 
and leads to a $2$-dim black hole metric with a dilaton background. 
The D-branes in this model are susy cycles and 
consistent with the analysis of conjugacy classes. 
The generalized metrics with U$(n)$ isometry is proposed 
and dynamics on them are realized by linear sigma models. 
We investigate D-branes of the linear sigma models and 
compare the results with those in the nonlinear sigma models.

}}

\end{center}

\newpage
\section{Introduction} 

Superstring theories have D-branes in open string sectors. 
However in general it is difficult to analyze 
properties of the branes on curved backgrounds. 
The CFT is a powerful tool to describe D-branes exactly,
but not convenient for understanding D-branes geometrically.
We want to know geometrical properties of the branes directly.
For this purpose, we study $N=2$ supersymmetric sigma models in this work. 

We investigate nonlinear sigma models with $2$-dimensional target space
as an example of curved space in the first part of this paper.
By considering A-,B-type supersymmetric boundary conditions on the worldsheet,
we get D1-branes for the A-type boundary and D0,D2-branes for the B-type boundary. 
We can construct boundary interaction terms added to the action. 
When the theory is required to be conformal invariant
the K\"ahler potential leads to a $2$-dim black hole metric with a dilaton background. 
Then the model is reduced to one of the Wess-Zumino-Witten models 
and the target space has the form of a cigar or a trumpet,
on which we give the geometric description of the D-branes.
The result is consistent with that in the analysis of conjugacy classes 
\cite{Bachas:2000fr,Elitzur:2001qd,Walton:2002db,Fotopoulos:2003vc,Ribault:2003ss}. 

The $2$-dim  black hole metric can be generalized to the K\"ahler metric with U$(n)$ isometry \cite{Kiritsis:pb},
which yields $2n$-dimensional spacetime.
D-branes are given by its submanifold. 
Moreover an effect of a dilaton on a point-like object is discussed. 

The discussion of the nonlinear sigma model
is a classical level analysis. 
Generally the sigma model action is modified by quantum corrections.
In Ref.\cite{Hori:2001ax}  the linear sigma  model equivalent to 
2-dim black hole at quantum level is proposed. 
Its generalization to  
a sigma model with the $2n$-dim target space
is thought to realize ${N=2}$ superconformal models with $c/3>1$ in the IR limit.
Therefore we consider such a linear sigma model 
and study its supersymmetric boundary conditions.
As a result we get D-branes consistent with those in the $2$-dim case.

The organization of this paper is as follows:
In section 2, we study                                                                                                                                                                                                                                                                                                                                                                                                                                                                                                                                                                                                                                                                                                                                                                                  
supersymmetric nonlinear sigma models with $2$-dim target space
and their boundary conditions.
Especially the model with the $2$-dim black hole metric is investigated in detail.
In section 3,
we analyze the generalized K\"ahler potential, which leads to $2n$-dim metric
and consider the relation between its submanifold and D-branes. 
In section 4
we consider the linear sigma model realizing dynamics on the above $2n$-dim target space
and examine its D-branes.
In section 5, we give summaries and conclusions.
In an appendix, we discuss a 
nonlinear sigma model with F-term.

\section{Nonlinear Sigma Model and 2-dim Black Hole}
\label{section_nonlinear}

In this section we study $N=(2,2)$ 
supersymmetric nonlinear sigma models on curved backgrounds.
The action for a world sheet $\Sigma$ without boundary is described
\beq
S &=& \frac{k}{4 \pi} \int_{\Sigma} d^2 z \Bigl[
\del\bar{\del}K \cdot(\del_{+}\phi\del_{-}\phib+
\del_{-}\phi\del_{+}\phib)\non
&& -i\chi^{+}\chi^{-}\left(\del_{+}\phi\cdot \frac{\del^2 \bar{\del}K}{\del\bar{\del}K}
-\del_{+}\phib \cdot \frac{\del \bar{\del}^2 K}{\del\bar{\del}K}\right)
-i\lambda^{+}\lambda^{-}\left(\del_{-}\phi\cdot \frac{\del^2 \bar{\del}K}{\del\bar{\del}K}
-\del_{-}\phib \cdot \frac{\del \bar{\del}^2K}{\del\bar{\del}K}\right)\non
&& +2\lambda^{+}\lambda^{-}\chi^{+}\chi^{-}
\frac{\del^2 \bar{\del}^2 K}{(\del\bar{\del}K)^2}  +i (\lambda^{+}\del_{-}\lambda^{-}
+\lambda^{-}\del_{-}\lambda^{+}
+\chi^{+}\del_{+}\chi^{-}
+\chi^{-}\del_{+}\chi^{+})\non
&&+\frac{1}{2}\ \del\bar{\del}K \cdot F \bar{F}-\bar{F} \frac{\del^2\bar{\del}K}{\del\bar{\del}K}  \cdot \lambda^{+}\chi^{+}
+F \frac{\del\bar{\del}^2K}{\del\bar{\del}K}  \cdot \lambda^{-}\chi^{-} \Bigr],
\label{action}
\eeq
where 
$k-2$ is a level and
$K(\phi,\phib)$ is a K\"ahler potential.
We use the following conventions:
\begin{eqnarray}
  &&	z=\xi_1+i\xi_2,\quad \zb=\xi_1-i\xi_2,\quad  d\xi_1 d \xi_2=\frac{i}2 d zd \zb = d^2z,\nn\\
&& \del_+ = \frac{\del}{\del z} = \frac12(\del_1-i\del_2),\quad 
\del_- = \frac{\del}{\del \zb} =\frac12(\del_1+i\del_2),\quad
\del = \frac{\del}{\del \phi}, \quad  \bar{\del} = \frac{\del}{\del \phib}. \nn
\end{eqnarray}
In the case of $K(\phi, \phib)= \phi \phib$, this model equals that on the flat background.
Setting
\begin{eqnarray}
&&\psi_+ = \sqrt{2} (\del\delb K )^{-1/2} \lamp, \quad \bar{\psi}_+ = \sqrt{2} (\del\delb K )^{-1/2} \lamm, \nn\\
&&\psi_- = \sqrt{2} (\del\delb K )^{-1/2} \chip, \quad \bar{\psi}_- = \sqrt{2} (\del\delb K )^{-1/2} \chim,
\label{def_fermion}
\end{eqnarray}
reduces the action to a canonical $N=(2,2)$ nonlinear supersymmetric sigma model.
Note that this redefinition of the fermions breaks down when $\del\bar{\del}K$ blows up.
Such a configuration of $\phi$, $\phib$ corresponds to a coordinate or curvature singularity
in the target space. 
The action (\ref{action}) 
is invariant under the following $N=2$ supersymmetry transformations;
\beq
&&\delta \phi =\sqrt{2} (\del \bar{\del} K)^{-1/2} (\epsilon_{+}\chi^{+}-\epsilon_{-}\lambda^{+}),  \quad 
\delta \phib =\sqrt{2}(\del \bar{\del}K)^{-1/2}(-\bar{\epsilon}_{+}\chi^{-}+\bar{\epsilon}_{-}\lambda^{-})\,,\non
&&\delta F= -2 
\sqrt{2} i \bar{\epsilon}_{+} \del_{-} \left( ( \del \bar{\del}K)^{-1/2}\lambda^{+} \right)
-2 \sqrt{2}i\bar{\epsilon}_{-}\del_{+} \left( (\del \bar{\del}K)^{-1/2}\chi^{+} \right)\,,\non
&&\delta \bar{F}=-2 \sqrt{2}i{\epsilon}_{+}\del_{-} \left( (\del \bar{\del}K)^{-1/2}\lambda^{-} \right)
-2\sqrt{2}i{\epsilon}_{-}\del_{+} \left( (\del \bar{\del}K)^{-1/2}\chi^{-} \right)\,,\non
&&\delta \lambda^{+}=\sqrt{2}(\del \bar{\del}K)^{1/2}\bar{\epsilon}_{-}
\left(i\del_{+}\phi -\frac{\del \bar{\del}^2 K}{2(\del \bar{\del}K)^{2}}\cdot 
\lambda^{+}\lambda^{-}\right)\non
&&\qquad +\frac{1}{\sqrt{2}}(\del \bar{\del}K)^{1/2}
\Biggl[
\epsilon_{+}\left(F+\frac{\del^2 \bar{\del}K}{(\del \bar{\del}K)^{2}}\cdot \chi^{+}\lambda^{+}\right)
-\frac{\del \bar{\del}^2 K}{(\del \bar{\del}K)^{2}}\cdot \bar{\epsilon}_{+}\chi^{-}\lambda^{+}
\Biggr]\,,\non
&&\delta \chi^{+}=\sqrt{2}(\del \bar{\del}K)^{1/2}\bar{\epsilon}_{+}
\left(-i\del_{-}\phi
+\frac{\del \bar{\del}^2 K}{2(\del \bar{\del}K)^{2}}\cdot \chi^{+}\chi^{-}\right)\non
&&\qquad +\frac{1}{\sqrt{2}}(\del \bar{\del}K)^{1/2}
\Biggl[
\epsilon_{-}\left(F+\frac{\del^2 \bar{\del}K}{(\del \bar{\del}K)^{2}} \cdot \chi^{+}\lambda^{+}\right)
-\frac{\del \bar{\del}^2 K}{(\del \bar{\del}K)^{2}}
\cdot \bar{\epsilon}_{-}\chi^{+}\lambda^{-}
\Biggr]\,,\non
&&\delta \lambda^{-}=\sqrt{2}(\del \bar{\del}K)^{1/2}\epsilon_{-}
\left(-i\del_{+}\phib
-\frac{\del^2 \bar{\del} K}{2(\del \bar{\del}K)^{2}}\cdot \lam^{+}\lam^{-}\right)\non
&&\qquad +\frac{1}{\sqrt{2}}(\del \bar{\del}K)^{1/2}
\Biggl[
\bar{\epsilon}_{+}\left(\bar{F}-\frac{\del \bar{\del}^2 K}{(\del \bar{\del}K)^{2}} \cdot \chi^{-}\lambda^{-}\right)
+\frac{\del^2 \bar{\del} K}{(\del \bar{\del}K)^{2}}
\cdot {\epsilon}_{+}\chi^{+}\lambda^{-}
\Biggr]\,,\non
&&\delta \chi^{-}=\sqrt{2}(\del \bar{\del}K)^{1/2}\epsilon_{+}
\left(i\del_{-}\phib
+\frac{\del^2 \bar{\del} K}{2(\del \bar{\del}K)^{2}}\cdot \chi^{+}\chi^{-}\right)\non
&&\qquad +\frac{1}{\sqrt{2}}(\del \bar{\del}K)^{1/2}
\Biggl[
\bar{\epsilon}_{-}\left(\bar{F} -\frac{\del \bar{\del}^2 K}{(\del \bar{\del}K)^{2}} \cdot \chi^{-} \lam^-\right)
+\frac{\del^2 \bar{\del} K}{(\del \bar{\del}K)^{2}}
\cdot {\epsilon}_{-}\chi^{-} \lam^+
\Biggr].
\label{susy}
\eeq
The singular point is invariant under (\ref{susy}) and thus forms a fixed point of the supersymmetry.
This model also has U(1) R-symmetry, whose charge is assigned $+1$ for $\chip$ and $\lamm$, and $-1$ for
$\chim$ and $\lamp$.

If $\Sigma$ has a boundary on $z=\zb$ ($\xi_2=0$), 
one should consider the model on the upper half plane 
$\Sigma = \{(\xi_1,\xi_2)\,|\,\xi_2\ge 0\}=\{z\,|\,{\rm Im} z\ge 0\}$.
Under the above susy transformation,  we can calculate the variation of $S$ as
\beq
\delta S &=& \frac{k}{4 \pi} \frac{1}{2}
\int_{\del \Sigma} d\xi_1 \Biggl[ 
i \sqrt{2}(\del \delb K)^{1/2}
\left( \epsbp \chim \delp \phi + \epsbm \lamm \delm \phi
- \epsp \chip \delp \phib -\epsm \lamp \delm \phib
\right) \nn \\
&& +\frac{1}{\sqrt{2}}(\del \delb K)^{1/2}
\Bigl[   
\epsm \left( F +2 \frac{\del^2 \delb K}{(\del \delb K)^{2}} \chip \lamp  \right) \chim
-\epsp \left( F +2 \frac{\del^2 \delb K}{(\del \delb K)^{2}} \chip \lamp \right) \lamm \nn \\
&&+\epsbm \left( \bar{F} -2 \frac{\del \delb^2 K}{(\del \delb K)^{2}} \chim \lamm \right) \chip
-\epsbp \left( \bar{F} -2 \frac{\del \delb^2 K}{(\del \delb K)^{2}} \chim \lamm \right) \lamp
\Bigr]
\Biggr].
\label{variation_action}
\eeq
Here we note that all the terms except those in the first line vanish due to the equations of motion of $F, \bar{F}$. 
If the $\del\bar{\del}K$ does not become zero nor blow up, this leads us to the 
discussion of the supersymmetric sigma models.
There exist two types of boundary conditions depending on how to mix left moving  and right moving parts of
the supersymmetry \cite{Ooguri:1996ck,Hori:2000ck,Govindarajan:2000ef,Hori:2000ic};
\begin{itemize}
\item  A-type boundary\, $\epsp = \epsbm = \eps, \quad \epsm = \epsbp = \epsb$\,;\\
an example of the boundary condition is
\begin{eqnarray}
	\lamm \pm \chip = \lamp \pm \chim =0, \nn \\
	\delm \phi \pm \delp \phib =0, \quad \delm \phib \pm \delp \phi =0, \nn
	\end{eqnarray}
\item B-type boundary\, $\epsp = -\epsm = \eps, \quad \epsbp = -\epsbm = \epsb$\,;\\
\begin{eqnarray}
	\lamm \pm \chim= \lamp \pm \chip =0, \nn \\
	\delm \phi \pm \delp \phi =0, \quad \delm \phib \pm \delp \phib =0.
\label{b-type_condition}
\end{eqnarray}
\end{itemize}
For the case of B-type $\epsp=-\epsm$ ($\epsbp=-\epsbm$),
we can introduce the following interaction terms on the boundary
\begin{eqnarray}
S_{bdy}=-\frac{k}{4 \pi} \frac12 \int_{\del \Sigma} d\xi_1 \left( \chip \lamm + \chim \lamp \right)\,,	
\label{boundary_int}
\end{eqnarray}
so that the supersymmetry variation vanishes 
$\delta \left( S + S_{bdy}\right)=0$. 
The ordinary variation of the fields on the boundary is proportional to 
\begin{eqnarray} 
&&\delta \phi \left(  
-\del \delb K \cdot \del_2 \phib + \frac{\del^2 \delb K}{\del \delb K}(\chip \chim -\lamp\lamm)
\right)\nn \\
&&+\delta \phib \left(  
-\del \delb K \cdot \del_2 \phi - \frac{\del \delb^2 K}{\del \delb K}(\chip \chim -\lamp\lamm)
\right)
\nn \\
&&+(\delta \chip -\delta \lamp) (\chim + \lamm)+ (\delta \chim - \delta \lamm) (\chip + \lamp),\nn
\label{henbun}
\end{eqnarray}
which yields boundary conditions on $z =\zb$;
\begin{eqnarray}
&& \chip \pm \lamp = \chim \pm \lamm=0,\nn\\
&&\delta \phi = \delta \phib =0\,\, \mbox{(Dirichlet) \,\, or}   
\quad
\del_2 \phi = \del_2 \phib =0\,\, \mbox{(Neumann)}.\nn
\end{eqnarray}
These boundary conditions are consistent with Eqs.(\ref{b-type_condition}).
The former condition for the bosonic part means the existence of D0-brane and the latter means that of D2-brane. 

We can choose K\"ahler metrics arbitrarily as a sigma model, but
the geometry is restricted when we embed the model in the 
superstring theories.
Here we require the theory to be conformal invariant, namely $\beta_{\mu \nu}^{G}=0$ 
\cite{Kiritsis:pb,Mandal:1991tz}.
The metric, Christoffel connections and Ricci tensors are given as:
\begin{eqnarray}
	&&ds^2 = \del \delb K \cdot d \phi d \phib, \quad K = K(\phi,\phib), \nn\\
      &&\Gamma^{\phi}_{\phi\phi}= \frac{\del^2\delb K}{\del\delb K},\quad 
      \Gamma^{\phib}_{\phib\phib}= \frac{\del\delb^2 K}{\del\delb K},\quad \mbox{(other connections)}=0,\nn\\
      &&R_{\phi \phib}= -\del \delb \log ( \del \delb K ), \quad R_{\phi \phi}=R_{\phib \phib}=0.\nn
\end{eqnarray}	
The condition for the $\beta$-function to vanish is
\begin{eqnarray}	
	R_{\mu\nu}= -2 \nabla_\mu \nabla_\nu \Phi, \nn
\end{eqnarray}
where $\Phi$ is a dilaton field, or
\begin{eqnarray}
      && 2 \del \delb \Phi = \del \delb \log (\del\delb K), \label{beta1}\\
	&&  \del^2 \Phi - \Gamma^{\phi}_{\phi\phi} \del \Phi =0, \label{beta2}\\
	&&  \delb^2 \Phi - \Gamma^{\phib}_{\phib\phib} \delb \Phi =0. \label{beta3}
\end{eqnarray}
From (\ref{beta1}) we can choose $2 \Phi = \log (\del\delb K)$.
With this relation  Eqs.(\ref{beta2}) and (\ref{beta3}) are rewritten into
$\del^2 e^{-2 \Phi}=0,\ \delb^2  e^{-2 \Phi}=0$.
Thus the following relation is obtained
\begin{eqnarray}
	e^{-2 \Phi} = (\del\delb K)^{-1}= A \phi \phib +B \phi +C \phib +D, \quad A, B, C, D:\mbox{constant}.\nn
\end{eqnarray}
If $A \neq 0$, we can set $B=C=0$ after a shift of $\phi, \phib$ and then get 
\begin{eqnarray*}
    	\del\delb K = \frac{1}{A \phi \phib +D}.	
\end{eqnarray*}
In the case of $D=0$ the model is reduced to trivial flat space geometry
after an appropriate transformation of $\phi$,$\phib$. 
So we look into the case of $AD\neq 0$.
We classify the metrics depending on signs of $A$ and $D$ in Table \ref{table:classify}.
\begin{table}[t]
\caption{Classification of the metric. The $\phi$ and $\phib$ are rescaled appropriately. 
The line element $ds^2$ is calculated by using the parameterization (\ref{sl2}) or (\ref{su2}). 
Each model is equivalent to a WZW model.}
\begin{center}
\renewcommand{\arraystretch}{1.3}
\begin{tabular}{c l l l l l}
	\hline
& $A,D$ & metric & $ds^2$ & WZW model 	\\
	\hline
(I) &$A>0\,,\,D>0$ & $\del\delb K=\frac{1}{\phi\bar{\phi}+1}$ & $dr^2+\tanh^2 r d \theta^2 $ &
SL(2;{\bf R})/U(1)${}_A$  \\ 
(II) &$A>0\,,\,D<0$ & $\del\delb K=\frac{1}{\phi\bar{\phi}-1}$ & $dr^2+\coth^2 r d \tilde{\theta}^2 $&
SL(2;{\bf R})/U(1)${}_V$  \\ 
(III) &$A<0\,,\,D>0$ & $\del\delb K=\frac{1}{-\phi\bar{\phi}+1}$ & $dr^2+\tan^2 r d \theta^2 $ &
  SU(2)/U(1)  \\
 & & &($dr^2+\cot^2 r d \tilde{\theta}^2$) &
\\	\hline
\end{tabular}
\end{center}
\label{table:classify}
\end{table}
When $K(\phi,\phib)$ is represented by the dilogarithm function
\begin{eqnarray}
K(\phi,\phib)=
-\mbox{Li}_2(\pm\phi\phib)=-\sum^{\infty}_{k=1}\frac{(\pm \phi \phib)^k}{k^2},\nn
\end{eqnarray}
this model has target space geometry with a 2-dim black hole metric\cite{Witten:1991yr,Nakatsu:1991pu}
\begin{eqnarray}
ds^2 = \frac{d \phi d \phib}{\phi \phib \pm 1}. \nn
\end{eqnarray}
The upper sign corresponds to the case (I) in Table \ref{table:classify}.
For this case if we set $\phi =u\,,\,\,\phib=-v$
the action (\ref{action}) leads to 
\begin{eqnarray}
S&=&\frac{k}{4\pi}\int d^2z
\Biggl[
-\frac{\partial u \bar{\partial} v +\partial v \bar{\partial} u}
{1-uv}
+i\chi^{+}\chi^{-}\frac{u \partial v -v \partial u}{1-uv}
+i\lambda^{+}\lambda^{-}
\frac{u\delb v -v\delb u}{1-uv}
\nn\\
&&-\frac{2}{1-uv}\chi^{+}\chi^{-}\lambda^{+}\lambda^{-} 
+i \left( 
\lamp \delm \lamm + \lamm \delm \lamp + \chip \delp \chim + \chim \delp \chip
\right)
\Biggr],
\label{action_axial}
\end{eqnarray}
after eliminating $F$, $\fb$ by the equations of motion.
This action is equal to one of the SL(2;{\bf R})/U(1) gauged Wess-Zumino-Witten models after gauge-fixing
and known as axial model.
On the other hand, the lower sign corresponds to the case (II).
For this case the same action as (\ref{action_axial}) 
in which $\lam^{\pm}$ are replaced with $\lam^{\mp}$ and $(u,v)$ with $(a,b)$ is obtained and
known as vector model.
This model is achieved by using a twisted chiral superfield instead of 
a chiral superfield in constructing a nonlinear sigma model and setting
$\phi =a\,,\,\,\phib=b$. 

\begin{figure}[b]
 \begin{center}
   \scalebox{0.55}{\input{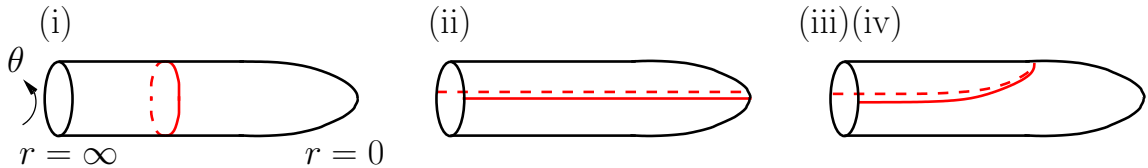}}
   \caption{The D-branes in the axial model. The target space metric of this model has the form of a cigar 
and  the horizon at $r=0$ (regular).}
   \label{figure:axial}
 \end{center}
 \end{figure}

\begin{figure}[t]
\begin{center}
   \scalebox{0.55}{\input{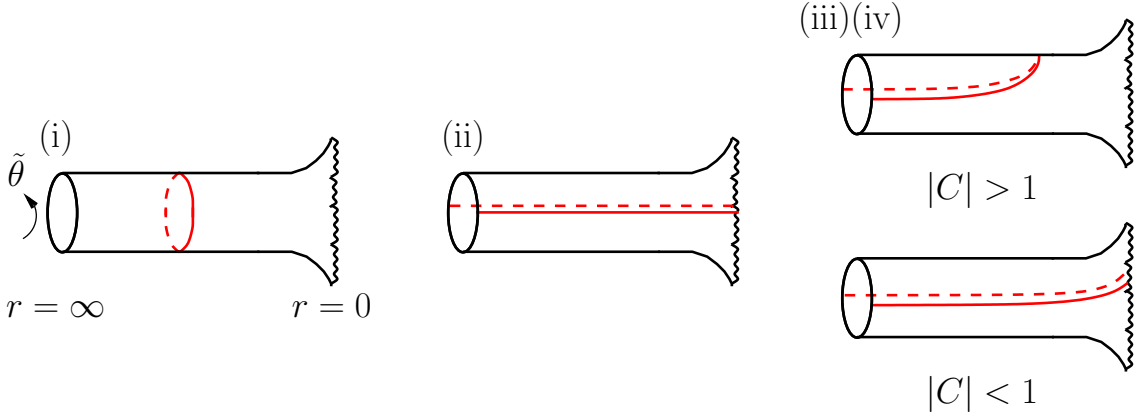}}
   \caption{The D-branes in the vector model. The target space metric of this model has the form of a trumpet 
and  a singularity at $r=0$. We set $C=\cosh r \cos \tilde{\theta}$ for (iii) or 
$C=\cosh r \sin \tilde{\theta}$ for (iv).}
   \label{figure:vector}
\end{center}
\end{figure}

To investigate geometrical shapes of the D-branes in the target space for the A-type boundary, we parameterize
\begin{eqnarray}
	\left( \begin{array}{cc}
	a & u \\
	-v & b
	\end{array}\right)
	=\left( \begin{array}{cc}
	e^{i \tilde{\theta}} \cosh r  &  e^{i \theta} \sinh r\\
	e^{-i \theta} \sinh r & e^{-i \tilde{\theta}} \cosh r
	\end{array}\right),
\label{sl2}
\end{eqnarray}
where $ab+uv=1,\ 0 \leq \theta<2 \pi,\ 0 \leq \tilde{\theta}<2 \pi$ and 
$0 < r < \infty$.
For the axial model
the integrand in Eq.(\ref{variation_action}) is rewritten into the formula proportional to
\begin{eqnarray}
&&\eps \left[  \lamm \delm (e^{i \theta} \sinh r) - \chip \delp (e^{-i\theta} \sinh r) \right]
+\epsb \left[  \chim \delp (e^{i \theta} \sinh r) - \lamp \delm (e^{-i\theta} \sinh r) \right] \nn \\	
&=&\frac12\eps \left[  (\lamm e^{i \theta} -\chip e^{-i \theta})( \cosh r \del_1 r -\sinh r \del_2 \theta)
 +i(\lamm e^{i \theta} +\chip e^{-i \theta})( \sinh r \del_1 \theta +\cosh r \del_2 r) \right]\nn\\
&+&\frac12\epsb \left[ i(\chim e^{i \theta} +\lamp e^{-i \theta})( \sinh r \del_1 \theta -\cosh r \del_2 r)
 +(\chim e^{i \theta} -\lamp e^{-i \theta})( \cosh r \del_1 r +\sinh r \del_2 \theta) \right]. \nn
\end{eqnarray}
We can set the boundary conditions in various ways;
\begin{eqnarray}
	\mbox{(i)}&& \chip e^{-i \theta}+\lamm e^{i \theta}= \chim e^{i \theta}+\lamp e^{-i \theta}=0,\quad 
	\del_1 r = \del_2 \theta=0, \nn\\
	\mbox{(ii)}&& \chip e^{-i \theta}-\lamm e^{i \theta}= \chim e^{i \theta}-\lamp e^{-i \theta}=0,\quad 
	\del_1 \theta = \del_2 r=0,\nn\\
	\mbox{(iii)}&& \chip + \lamm = \chim + \lamp=0, 
	\quad \del_{1} (\cos \theta \sinh r) =\del_{2} (\sin \theta \sinh r) =0,\nn\\
	\mbox{(iv)}&& \chip - \lamm = \chim - \lamp=0, 
	\quad \del_{1} (\sin \theta \sinh r) =\del_{2} (\cos \theta \sinh r) =0\nn,
\end{eqnarray}
and illustrate the D1-branes expressed by these in Fig.\ref{figure:axial}.
Similarly for the vector model we can set the boundary conditions;
\begin{eqnarray}
	\mbox{(i)}&& \chip e^{-i \tilde{\theta}}-\lamp e^{i \tilde{\theta}}
	= \chim e^{i \tilde{\theta}}-\lamm e^{-i \tilde{\theta}}=0, \quad
	\del_1 r = \del_2 \tilde{\theta}=0, \nn\\
	\mbox{(ii)}&& \chip e^{-i \tilde{\theta}}+\lamp e^{i \tilde{\theta}}
	= \chim e^{i \tilde{\theta}}+\lamm e^{-i \tilde{\theta}}=0,\quad 
	\del_1 \tilde{\theta} = \del_2 r=0,\nn\\
	\mbox{(iii)}&& \chip - \lamp = \chim - \lamm=0, \quad 
	\del_{1} (\cos \tilde{\theta} \cosh r) =\del_{2} (\sin \tilde{\theta} \cosh r) =0, \nn\\
	\mbox{(iv)}&& \chip + \lamp = \chim + \lamm=0, \quad 
	\del_{1} (\sin \tilde{\theta} \cosh r) =\del_{2} (\cos \tilde{\theta} \cosh r) =0,\nn
	\end{eqnarray}
and illustrate them in Fig.\ref{figure:vector}.
These results are consistent with the semiclassical descriptions of 
D-branes in Refs.\cite{Bachas:2000fr,Fotopoulos:2003vc},
where D-branes in the SL(2;{\bf R})/U(1) gauged WZW model are analyzed 
by considering conjugacy classes of SL(2;{\bf R}).

\begin{figure}[h]
 \begin{center}
   \scalebox{0.5}{\input{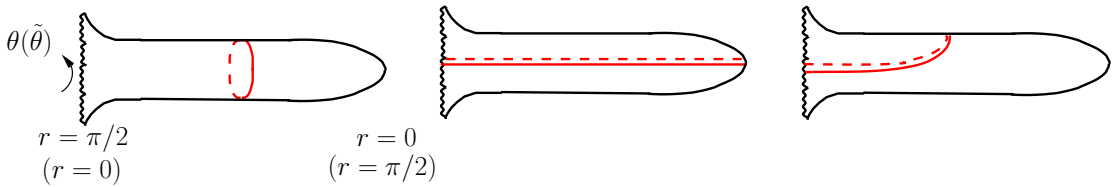}}
   \caption{The D-branes in the axial (vector) model. The target space has a singularity at $r=\pi/2$ ($r=0$)
and the horizon at $r=0$ ($r=\pi/2$). }
   \label{figure:su2}
 \end{center}
 \end{figure}

To investigate D-branes inside the horizon, $r$ needs to be analytically continued. 
The model corresponds to the case (III) in Table \ref{table:classify}   
and leads to an SU(2)/U(1) WZW model.
In this case we parameterize 
\begin{eqnarray}
&&	\left( \begin{array}{cc}
	a & u \\
	-v & b
	\end{array}\right)
	=\left( \begin{array}{cc}
	e^{i \tilde{\theta}} \cos r  &  i e^{i \theta} \sin r\\
	i e^{-i \theta} \sin r & e^{-i \tilde{\theta}} \cos r
	\end{array}\right), 
\label{su2}
\end{eqnarray}
where $0 \leq \theta ,\tilde{\theta} < 2 \pi$ and $0< r < \pi/2$.
By setting $(\phi,\phib)=(u,v)$ for the axial model or $(a,b)$ for the vector model,
several boundary conditions are obtained by the same procedure as above (see Fig.\ref{figure:su2}).
	
In the above discussion, we neglect a dilaton.
Its effect makes some of D-branes unstable and
several D-branes in these figures disappear.
	

\section{Generalized Metrics}
\label{section_generalize}

In the previous section, we studied brane configurations 
in the $2$-dimensional black hole. 
In this section, we study 
$2n$-dimensional metrics with U($n$) isometries\cite{Kiritsis:pb}, 
whose spacetimes are 
the generalization of the $2$-dimensional black hole.

\begin{figure}[b]
 \begin{center}
   \scalebox{0.55}{\input{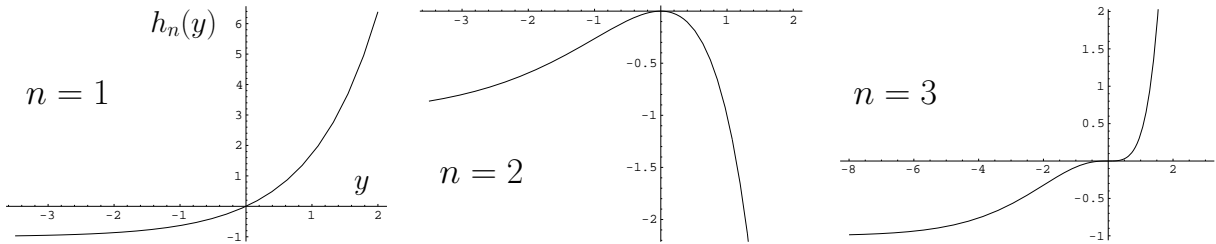}}
   \caption{The function $h_n(y)$ for $n=1,2,3$}
   \label{figure:n=123}
 \end{center}
 \end{figure}

The set of holomorphic coordinates is 
given by 
$(u_1,u_2,\cdots ,u_n)$ and K\"ahler potential of the space 
is a function of $x=\sum^n_{i=1}|u_i|^2$
 on the appropriate local coordinate patch.
Before writing down the metric, we parameterize $u_i$'s by 
$x$, $\theta$ and $w_i$'s 
\ba
\so u_i=x^{1/2} F^{-1/2} e^{i\theta/n}
\cdot w_i
\qquad (i=1,2,\cdots ,n\,;\,w_n=1)\,,\non
\so F=1+\sum^{n-1}_{i=1}|w_i|^2\,.\nom
\ea
This expression is valid for $x\neq 0$ and
 $w_i$'s are coordinates of CP${}^{n-1}$.
(The case of $x=0$ corresponds to the point $u_i=0$ ($i=1, 2,\cdots ,n$).)
Let us introduce a new coordinate $Y$ through an equation
\ba
\so Bx^n=h_n(\frac{2 n Y}{k}), \nn\\
&& h_n(y)=\frac{(-1)^{n-1}}{(n-1)!}\int^y_0dt\,t^{n-1}e^t=
-1+e^y \sum^{n-1}_{m=0}\frac{(-y)^m}{m!} \quad (B\,:\,\mbox{constant})\,.\nom
\ea
We plot the function $h_n(y)$ for $n=1,2,3$ in Fig. \ref{figure:n=123}.
The $h_n(y)$ becomes zero only at $y=0$,
while $h_n(y)$ approaches $-1$ in the $y \rightarrow -\infty$ limit.
If $n$ is even, $h_n(y)\leq 0$ and 
$B$ must be a negative constant. 
If $n$ is odd, $h_n(y)$ increases monotonously and 
$B$ can be either positive or negative. 
Since the associated K\"ahler potential $K=K(x)$ is defined by a solution of
$x\frac{dK}{dx}=Y(x)$, the metric is expressed as 
\ba
\so ds^2=\frac{g_n(Y)}{2}dY^2+\frac{2}{n^2g_n(Y)}
(d\theta -nA)^2
+2Yds_{FS}^2\,,\non
\so ds_{FS}^2=
F^{-1}\sum_{i,j=1}^{n-1}
\left(
\delta_{ij}-F^{-1}\bar{w}_iw_j
\right)dw_id\bar{w}_j\,,\non
\so \Phi =-\frac{nY}{k}+ C \quad (C\,:\,\mbox{constant}), \quad 
e^{2C}=B\cdot n!(-1)^{n-1}\left( \frac{2 n}{k} \right)^{-n},\nn
\ea
where we define 
\begin{eqnarray*}
&&g_n(Y)=\frac{2n}{k}\left\{f_n(\frac{2nY}{k})\right\}^{-1},\quad
f_n(Y)=nY^{1-n}e^{-Y}\int^Y_0dt\,t^{n-1}e^t,\quad \nn\\
&&A=-\frac{i}{2}F^{-1}\sum^{n-1}_{i=1}
(w_id\bar{w}_i-\bar{w}_idw_i)\,.\non
\end{eqnarray*}
The line element $ds_{FS}^2$ represents 
the Fubini-Study metric of CP${}^{n-1}$ and $A$ is a connection 
$1$-form on CP${}^{n-1}$.
The K\"ahler geometry is a fibered space over the 
CP${}^{n-1}$. 
The dilaton $\Phi$
is introduced so that 
the one-loop beta function becomes zero.
The case of $n=1$ corresponds to the 2-dim black hole treated in Section \ref{section_nonlinear}. 

At $Y=0$ the function $f_n(Y)$ is zero and $g_n(Y)$ blows up.
However the scalar curvature 
\begin{eqnarray*}
	R=\frac{4 n}{k}\Bigl[ n-f_n \left( \frac{2 nY}{k}\right) \Bigr]
\end{eqnarray*} 
is equal to $4 n^2/k$ and the geometry is regular there. 
Thus $Y=0$ is just a coordinate singularity of this system. 
In fact $Y=0$ corresponds to the tip of the 
cigar of the 2-dim black hole in the case of $n=1$.
In the $Y\rightarrow \infty$ limit, 
$R$ approaches zero,
while $Y \rightarrow -\infty$ limit $R$ blows up.
\begin{figure}[h]
 \begin{center}
   \scalebox{0.55}{\input{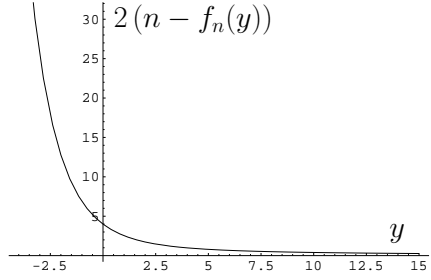}}
   \caption{The scalar curvature for n=2. The function $2(n-f_n(y))$ is plotted. 
We also obtain similar curves for the other values of $n$.}
   \label{figure:curvature}
 \end{center}
 \end{figure}

Here we consider the $N=2$ supersymmetric 
nonlinear sigma model on this curved background.
The action is given as
\ba
&& S=\int_{\Sigma}d^2z\,
\Biggl[
2g_{i\bar{j}}(
\del_{+}\phi^{i}\del_{-}\bar{\phi}^{\bar{j}}
+\del_{-}\phi^{i}\del_{+}\bar{\phi}^{\bar{j}})
+\frac{i}{2}g_{i\bar{j}}\bar{\psi}^{\bar{j}}_{-}
(D_0+D_1)\psi^i_{-}
+\frac{i}{2}g_{i\bar{j}} {\psi}^{i}_{-}
(D_0+D_1)\bar{\psi}^{\bar{j}}_{-}\non
&&\qquad +\frac{i}{2}g_{i\bar{j}}\bar{\psi}^{\bar{j}}_{+}
(D_0-D_1)\psi^i_{+}
+\frac{i}{2}g_{i\bar{j}} {\psi}^{i}_{+}
(D_0-D_1)\bar{\psi}^{\bar{j}}_{+}
-R_{i\bar{k}j\bar{\ell}}\psi^i_{+}\psi^j_{-}
\bar{\psi}^{\bar{k}}_{+}\bar{\psi}^{\bar{\ell}}_{-}
\Biggr]\,,\nom
\ea
where $g_{i{\bar{j}}}$ is the K\"ahler metric realizing the above geometry.
Supersymmetric transformation is 
defined by
\ba
&&\delta\phi^i=\ep_{+}\psi^i_{-}-\ep_{-}\psi^i_{+}\,,\non
&&
\delta\psi^i_{+}=
2i\bep_{-}\del_{+}\phi^i
+\ep_{+}\Gamma^i{}_{jk}\psi^j_{+}\psi^k_{-}\,,\non
&&
\delta\psi^i_{-}=
-2i\bep_{+}\del_{-}\phi^i
+\ep_{-}\Gamma^i{}_{jk}\psi^j_{+}\psi^k_{-}\,.\nom
\ea
Under this transformation, 
the variation of the action is calculated 
\ba
&&\delta S=\int_{\del\Sigma}d\xi^1\,
\Bigl[
-\ep_{+}g_{i\bar{j}}\psi^i_{-}\del_{+}\bar{\phi}^{\bar{j}}
-\ep_{-}g_{i\bar{j}}\psi^i_{+}\del_{-}\bar{\phi}^{\bar{j}}
+\bep_{+}g_{i\bar{j}}\bar{\psi}^{\bar{j}}_{-}\del_{+}\phi^i
+\bep_{-}g_{i\bar{j}}\bar{\psi}^{\bar{j}}_{+}\del_{-}\phi^i
\Bigr]\,.\nom
\ea
Let us examine the A-type boundary  
with $\ep_{+}=\bep_{-}=\ep$ and $\bep_{+}=\ep_{-}=\bep $.
By decomposing the fields $\phi_i$'s into 
angular parts $\varphi_i$ and radial parts $|u_i|$ like
$\phi_i =|u_i|e^{i\varphi_i}$,  
the variation is rewritten into
\ba
&&\delta S=\frac{1}{2}\int_{\del\Sigma}d\xi^1\,
\Bigl[
\ep \bigl\{
(-\psi^i_{-}e^{-i\varphi_i}+\bar{\psi}^{\bar{i}}_{+}e^{i\varphi_i})
(K'\delta_{ij}+\frac{1}{2}K''|u_i||u_j|)
(\del_1|u_j|-|u_j|\del_2\varphi_j)\non
&&\qquad 
+i(\psi^i_{-}e^{-i\varphi_i}+\bar{\psi}^{\bar{i}}_{+}e^{i\varphi_i})
(K'\delta_{ij}+\frac{1}{2}K''|u_i||u_j|)
(\del_2|u_j|+|u_j|\del_1\varphi_j)
\bigr\}\non
&&\qquad 
+\bep \bigl\{
(-\psi^i_{+}e^{-i\varphi_i}+\bar{\psi}^{\bar{i}}_{-}e^{i\varphi_i})
(K'\delta_{ij}+\frac{1}{2}K''|u_i||u_j|)
(\del_1|u_j|+|u_j|\del_2\varphi_j)\non
&&\qquad 
-i(\psi^i_{+}e^{-i\varphi_i}+\bar{\psi}^{\bar{i}}_{-}e^{i\varphi_i})
(K'\delta_{ij}+\frac{1}{2}K''|u_i||u_j|)
(\del_2|u_j|-|u_j|\del_1\varphi_j)
\bigr\}
\Bigr]\,.\nom
\ea
{}From this formula, we can read 
possible boundary conditions
\ba
&&\psi^i_{-}e^{-i\varphi_i}+\bar{\psi}^{\bar{i}}_{+}e^{i\varphi_i}=0\,,\,\,
\del_1|u_i|=0\,,\,\,\del_2\varphi_i=0\qquad 
(i=1,2,\cdots ,n)\,,\non
&&\psi^i_{-}e^{-i\varphi_i}-\bar{\psi}^{\bar{i}}_{+}e^{i\varphi_i}=0\,,\,\,
\del_2|u_i|=0\,,\,\,\del_1\varphi_i=0\qquad 
(i=1,2,\cdots ,n)\,.\nom
\ea
The former condition fixes $|u_i|$'s on the boundary.
The latter condition fixes $\varphi_i$'s and
this configuration corresponds to the Lagrangian submanifold.
We look into K\"ahler form $J$; 
\ba
\so J=\sqrt{-1}K_{u_iu_{\bar{j}}}du_i\wedge du_{\bar{j}}
=\frac{dK(x)}{dx}\sum_{i=1}^n d|u_i|^2\wedge d\hat{\varphi}_i
+\frac{d^2K(x)}{dx^2}\sum_{i,j=1}^n|u_j|^2d|u_i|^2\wedge 
d\hat{\varphi}_j\,.\non
\so \hat{\varphi}_i=\mbox{arg}\,u_i\qquad (i=1,2,\cdots ,n)\,.\nom
\ea
From this formula it turns out that  
the radial parts $|u_i|$'s and phase parts $\hat{\varphi}_i$'s  
are complementary one another. 
So we can consider Lagrangian submanifold specified 
by $\{\hat{\varphi}_i=\mbox{constant}\}$ (more precisely, 
linear combinations of $\hat{\varphi}_i$'s are 
constants on the appropriate coordinate patch). 
The set of coordinates is labelled by $|u_i|$'s. 
Thus pullback of the K\"ahler form $J$ vanish 
on this submanifold. 


To understand the structure of this fibered space, 
we use the coordinates $w_i$'s and $x$ instead of $u_i$'s.
Then
\ba
J&=&d\Bigl[ Y\cdot \left( \frac1n d \theta -A \right)\Bigr] \nn\\
&=& dY\wedge (\frac{1}{n}d\theta -A)
+YF^{-1}dF\wedge A
+YF^{-1}\sum_{i=1}^{n-1}d|w_i|^2\wedge d\varphi_i \,,\non 
A&=&-F^{-1}\sum_{i=1}^{n-1}|w_i|^2d\varphi_i\,, \qquad
 \varphi_i=\mbox{arg}\,w_i\,\qquad (i=1,2,\cdots ,n-1)\,.\nom
\ea
Note that $dY$ and $dF$ do not depend on 
$\theta$ nor $\varphi_i$'s. 
When one considers a subspace 
${\cal M}$ parameterized by 
the coordinates $Y$ and $|w_i|^2$ $(i=1,2,\cdots ,n-1)$, 
the pullback of this K\"ahler form vanish 
on ${\cal M}$. 
It is nothing else but the condition 
for the Lagrangian submanifold. 
It is a kind of minimal volume surfaces with 
middle dimension. 
The entire space can be considered to be a fibered space 
over this submanifold. Its fiber is 
$n$-dimensional torus parameterized by $\theta$ and $\varphi_i$ 
$(i=1,2,\cdots ,n-1)$.


Generally 2-dim $N=(2,2)$ supersymmetric sigma models have  
A-,B-type supersymmetric boundary conditions\cite{Ooguri:1996ck,Hori:2000ck}.
The A-type boundary conditions correspond to 
middle dimensional cycles, which
are Lagrangian submanifolds. 
For the discussed K\"ahler metric, (subsets of) ${\cal M}$ 
belongs to this kind of submanifolds.
It corresponds to nonconpact curves described by 
$\theta(\tilde{\theta}) =\mbox{constant}$ (see Fig.\ref{figure:axial}(ii))
for the $2$-dim black hole case ($n=1$).

For the B-type, 
the variation of the action can be written
with $\ep_{+}=-\ep_{-}=\ep$ and $\bep_{+}=-\bep_{-}=\bep$
\ba
&&\delta S=\frac{1}{2}
\int_{\del\Sigma}\xi^1\,
\Bigl[
\ep \left\{
g_{i\bar{j}}(-\psi^i_{-}+\psi^i_{+})\del_1\bar{\phi}^{\bar{j}}
+ig_{i\bar{j}}(\psi^i_{-}+\psi^i_{+})\del_2\bar{\phi}^{\bar{j}}
\right\}\non
&&\qquad 
+\bep \left\{
g_{i\bar{j}}(\bar{\psi}^{\bar{j}}_{-}
-\bar{\psi}^{\bar{j}}_{+})\del_1{\phi}^{i}
-ig_{i\bar{j}}(\bar{\psi}^{\bar{j}}_{-}
+\bar{\psi}^{\bar{j}}_{+})\del_2{\phi}^{i}
\right\}
\Bigr]\,.\nom
\ea
From this formula, it turns out that 
the B-type boundary corresponds 
to holomorphic cycles. 
Especially there are $0$-dim point-like objects and 
objects spreading over $2n$-dim spacetime.

For this K\"ahler geometry, 
there exists a dilaton $\Phi$ and 
the brane tension is proportional to 
$e^{nY/k}\cdot e^{-C}$. 
For odd $n$, $B$ must be positive and 
tension is minimized at $Y=0$ or $x=0$. 
In this case, the point-like object should be 
 localized at $x=0$. 
For even $n$, $B$ must be negative. 
However there are two branches of 
the coordinate transformation from $Y$ to $x$.
If we start at $Y>0$ and increase $x$, 
$Y$ increases and the tension gets greater. 
In order to minimize the tension, 
we should be at $Y=0$ ($x=0$) on this branch. 
So the position of this point-like object should be $Y=0$ ($x=0$). 
Meanwhile if we start at $Y<0$ and 
increase $x$, $Y$ decreases and 
$Bx^n$ approaches $-1$. 
Then the tension becomes zero.
On the other hand
the geometry blows up because of the curvature singularity. 
The description of geometry breaks down in this limit. 
Accordingly, if we consider $Y\geq 0$ and $x\geq 0$ 
branch, we obtain the result that a point-like object is localized at $Y=0$ ($x=0$).
For the case of $n=1$,  
such an object is the D$0$-brane localized on the tip of the cigar.

\newpage
\section{Linear Sigma Model}

In section \ref{section_nonlinear}, we discussed the nonlinear sigma model and
particularly investigated the model whose geometry is 2-dim black hole. 
As a result two types of supersymmetric boundary conditions are obtained, 
but the analysis is at classical level. 
In general the sigma model action is modified 
by quantum corrections.
In Ref.\cite{Hori:2001ax}  
the linear sigma  model equivalent to 
2-dim black hole at quantum level is proposed. 
Its generalization to  
a sigma model with the target space discussed in Section \ref{section_generalize} 
is believed to 
realize ${N=2}$ superconformal models with $c/3>1$ 
in the IR limit.
So let us study supersymmetric boundary conditions 
of such a linear sigma model. 

We use conventions 
for 2-dim worldsheet coordinates $(x^0,x^1)$
and derivatives
\ba
\so ds^2=-(dx^0)^2+(dx^1)^2\,,\,\,
\eta_{\alpha\beta}=(-1, +1)\,,\non
\so x^{\pm}=x^0\pm x^1\,,\,\,
 \del_{\pm}=\frac{1}{2}(\del_0\pm \del_1)\,.\nom
\ea
At first, we consider 
a linear sigma model containing 
chiral superfields $\Phi$, $P$ and 
a vector superfield $V$
\begin{eqnarray*}
\so \Phi =\phi (y)+\theta^{+}\psi_{+}(y)+\theta^{-}\psi_{-}(y)
+\theta^{+}\theta^{-}F(y)\,,\non
\so P =p (y)+\theta^{+}\chi_{+}(y)+\theta^{-}\chi_{-}(y)
+\theta^{+}\theta^{-}F_P(y)\,,\non
\so V=\theta^{-}\bar{\theta}^{-}(v_0-v_1)+\theta^{+}\bar{\theta}^{+}
(v_0+v_1)-\theta^{-}\bar{\theta}^{+}\sigma -\theta^{+}\bar{\theta}^{-}
\bar{\sigma}\non
\so \qquad +i\theta^{-}\theta^{+}(\bar{\theta}^{-}\bar{\lambda}_{-}
+\bar{\theta}^{+}\bar{\lambda}_{+}) 
+i\bar{\theta}^{+}\bar{\theta}^{-}(\theta^{-}\lambda_{-}
+\theta^{+}\lambda_{+})+\theta^{-}\theta^{+}\bar{\theta}^{+}
\bar{\theta}^{-}D\,,
\end{eqnarray*}
where we take Wess-Zumino gauge for $V$.
We use the following conventions
\begin{eqnarray*}
\so v_{01}=\del_0v_1-\del_0v_1\,,\,\,
 y^{\pm}=x^{\pm}-i\theta^{\pm}\bar{\theta}^{\pm}\,,\,\,
 \tilde{y}^{\pm}=x^{\pm}\mp i\theta^{\pm}\bar{\theta}^{\pm}\,,\non
\so D_{\pm}=+\frac{\del}{\del\theta^{\pm}}
-i\bar{\theta}^{\pm}\del_{\pm}\,,\qquad 
\bar{D}_{\pm}=-\frac{\del}{\del\bar{\theta}^{\pm}}
+i{\theta}^{\pm}\del_{\pm}\,,
\end{eqnarray*}
then the field strength of $V$ is given by $\Sigma$
\begin{eqnarray*}
\so \Sigma =\bar{D}_{+}D_{-}V
=\sigma (\tilde{y})+i\theta^{+}\bar{\lambda}_{+}
(\tilde{y})-i\bar{\theta}^{-}\lambda_{-}(\tilde{y})
+\theta^{+}\bar{\theta}^{-}(D-iv_{01})(\tilde{y})\,.\non
\end{eqnarray*}
The action $S$ of the linear sigma model \cite{Hori:2001ax,Witten:1993yc} is
\begin{eqnarray*}
S&=&\frac{1}{2\pi}\int dx^0dx^1\,( L_K+L_P+L_g+L_{\Sigma})\,,\non
\end{eqnarray*}
and each Lagrangian density is expressed by the component fields;
\begin{eqnarray*}
L_K&=&\int d^4\theta \,\bar{\Phi}e^V\Phi\nn\\
&=&D_0\bar{\phi}D_0\phi -D_1\bar{\phi}D_1\phi 
+\frac{i}{2}\bar{\psi}_{-}(D_0+D_1){\psi}_{-}
+\frac{i}{2}{\psi}_{-}(D_0+D_1)\bar{\psi}_{-}\non
&&+\frac{i}{2}\bar{\psi}_{+}(D_0-D_1){\psi}_{+}
+\frac{i}{2}{\psi}_{+}(D_0-D_1)\bar{\psi}_{+}
+D|\phi|^2+|F|^2-|\sigma |^2|\phi |^2\non
&&-\sigma \bar{\psi}_{-}\psi_{+}
-\bar{\sigma}\bar{\psi}_{+}\psi_{-}
-i\bar{\phi}(\lambda_{-}\psi_{+}-\lambda_{+}\psi_{-})
-i{\phi}(\bar{\psi}_{-}\bar{\lambda}_{+}-\bar{\psi}_{+}\bar{\lambda}_{-})
\,,\non
L_P&=&\frac{k}{4}\int d^4\theta\, (P+\bar{P}+V)^2\,\nn\\
&=&\frac{k}{2}
\Bigl(
D_0\bar{p}D_0p-D_1\bar{p}D_1p
+i\chi_{+}\del_{-}\bar{\chi}_{+}
+i\bar{\chi}_{+}\del_{-}{\chi}_{+}\non
 &&+i\chi_{-}\del_{+}\bar{\chi}_{-}
+i\bar{\chi}_{-}\del_{+}{\chi}_{-}
+i\chi_{+}\lambda_{-}-i\chi_{-}\lambda_{+}\non
&&+i\bar{\chi}_{+}\bar{\lambda}_{-}-i\bar{\chi}_{-}\bar{\lambda}_{+}
+|F_P|^2-|\sigma |^2+D(p+\bar{p})
\Bigr)\,,\non
L_g&=&-\frac{1}{2e^2}\int d^4\theta\, \bar{\Sigma}\Sigma\,\nn\\
&=&\frac{1}{2e^2}
\Bigl(
\del_0\bar{\sigma}\del_0\sigma
-\del_1\bar{\sigma}\del_1\sigma
+i\lambda_{+}\del_{-}\bar{\lambda}_{+}
+i\bar{\lambda}_{+}\del_{-}{\lambda}_{+}
+i\lambda_{-}\del_{+}\bar{\lambda}_{-}\non  
&&+i\bar{\lambda}_{-}\del_{+}{\lambda}_{-}
+D^2+v_{01}^2 \Bigr)\,,\non
L_{\Sigma}&=&\frac{1}{2}\int d^2\tilde{\theta}\,
(-t)\Sigma +\frac{1}{2}\int d^2\bar{\tilde{\theta}}
(-\bar{t})\bar{\Sigma}\,
 =-rD+\theta v_{01}\,, \nn\\
\end{eqnarray*}
where we used the following formulae
\begin{eqnarray*}
\so D_a\phi =(\del_a+iv_a)\phi\,,\quad D_a\bar{\phi}=
(\del_a-iv_a)\bar{\phi}\,,\non
\so D_ap=\del_ap+iv_a\,,\quad D_a\bar{p}=\del_a\bar{p}-iv_a 
\quad (a=0,1),\non
&&t=r-i\theta\,.\nom
\end{eqnarray*}
The system is invariant under the gauge transformation;
\ba
V\rightarrow V-i\Lambda +i\bar{\Lambda}\,,\quad
\Phi\rightarrow e^{i\Lambda}\Phi\,,\quad
P\rightarrow P+i\Lambda\,.\label{gauge}
\ea
The $\Lambda$ $(\bar{\Lambda})$ is a 
chiral (an anti-chiral) superfield.
This action also has $N=(2,2)$ supersymmetry 
on $\Sigma$
\begin{eqnarray*}
&&\de \phi = \ep_{+}\ps_{-}-\ep_{-}\ps_{+}\,,\non
&&\de \ps_{+} = +i(D_0+D_1)\phi\cdot \bep_{-}
+\ep_{+}F-\phi\bar{\sigma}\bep_{+}\,,\quad
\de \ps_{-} = -i(D_0-D_1)\phi\cdot \bep_{+}
+\ep_{-}F+\phi {\sigma}\bep_{-}\,,\non
&&\de F = -i \{\bep_{+}(D_0-D_1)\ps_{+}
+\bep_{-}(D_0+D_1)\ps_{-}\}
+\bep_{+}\bar{\sigma}\ps_{-}+\bep_{-}\sigma\ps_{+}
+i\phi (\bep_{-}\bar{\lambda}_{+}-\bep_{+}\bar{\lambda}_{-})\,,\non
\end{eqnarray*}
\begin{eqnarray*}
&& \de p = \ep_{+}\chi_{-}-\ep_{-}\chi_{+}\,,\non
&&\de \chi_{+} = +i(D_0+D_1)p\cdot \bep_{-}
+\ep_{+}F_P-\bar{\sigma}\bep_{+}\,,\quad
\de \chi_{-} = -i(D_0-D_1)p\cdot \bep_{+}
+\ep_{-}F_P+ {\sigma}\bep_{-}\,,\non
&&\de F_P = -2i (\bep_{+}\del_{-}\chi_{+}
+\bep_{-}\del_{+}\chi_{-})
+i  (\bep_{-}\bar{\lambda}_{+}-\bep_{+}\bar{\lambda}_{-})\,,\non
\end{eqnarray*}
\begin{eqnarray*}
&&\de v_0=\frac{i}{2}(\bep_{+}\lambda_{+}+\bep_{-}\lambda_{-}
+\ep_{+}\bar{\lambda}_{+}+\ep_{-}\bar{\lambda}_{-})\,,\quad
\de v_1=\frac{i}{2}(\bep_{+}\lambda_{+}-\bep_{-}\lambda_{-}
+\ep_{+}\bar{\lambda}_{+}-\ep_{-}\bar{\lambda}_{-})\,,\non
&&\de \sigma = -i(\bep_{+}\lambda_{-}
+\ep_{-}\bar{\lambda}_{+})\,,\quad
\de D = -\bep_{+}\del_{-}\lambda_{+}
-\bep_{-}\del_{+}\lambda_{-}
+\ep_{+}\del_{-}\bar{\lambda}_{+}
+\ep_{-}\del_{+}\bar{\lambda}_{-}\,,\non
&&\de \lambda_{+} = +i\ep_{+}(D+iv_{01})+2\del_{+}
\bar{\sigma}\cdot \ep_{-}\,,\quad
\de \lambda_{-} = +i\ep_{-}(D-iv_{01})+2\del_{-}
{\sigma}\cdot \ep_{+}\,.\nom
\end{eqnarray*}
To compare with the nonlinear sigma model, 
we examine the potential terms $U$
of scalar fields $\phi$, $p$ and $\sigma$
\ba
\so U=D|\phi |^2+|F|^2-|\sigma |^2|\phi |^2-\sigma \bar{\psi}_{-}
\psi_{+}-\bar{\sigma}\bar{\psi}_{+}\psi_{-}\non
\so \qquad +\frac{k}{2}D(p+\bar{p})+\frac{k}{2}|F_P|^2-\frac{k}{2}
|\sigma |^2+\frac{1}{2e^2}(v_{01}^2+D^2)-rD+\theta v_{01}\,.\nom
\ea
The equations of motion leads to
\ba
\so F=F_P=0\,,\,\,v_{01}=-\theta e^2\,,\non
\so -\frac{D}{e^2}=|\phi |^2+\frac{k}{2}(p+\bar{p})-r\,.\nom
\ea
The target space metric of the model is originally flat
because this system is a  
linear sigma model.
In the $e^2 \rightarrow \infty$ limit, 
we get the equations
\ba
\so \sigma =-\left(|\phi |^2+\frac{k}{2}\right)^{-1}
\bar{\psi}_{+}\psi_{-}\,,\,\,
\bar{\sigma}=-\left(|\phi |^2+\frac{k}{2}\right)^{-1}
\bar{\psi}_{-}\psi_{+}\,.\nom
\ea
After one eliminates the field $\sigma $, 
the four-fermion interaction term is induced in U
\ba
\so U=
\left(|\phi |^2+\frac{k}{2}\right)^{-1}
\bar{\psi}_{+}\psi_{-}\bar{\psi}_{-}\psi_{+}
+\cdots \,,\nom
\ea
which corresponds 
to that of the nonlinear sigma model.
It encodes information on curvature of the resulting target space geometry.

Now we consider susy conditions on the boundary $\del\Sigma 
=(x^0,0)$.
The susy transformation of $L=L_K+L_P+L_g+L_{\Sigma}$ is 
\ba
&&\delta L=\frac{1}{2}\del_1{\cal Q}+\del_0({\cdots})\,,\non
\so {\cal Q}=
\ep_{+}\Bigl[
-\psi_{-}(D_0+D_1)\bar{\phi}+i\sigma\bar{\phi}\psi_{+}
+i\bar{\psi}_{+}F+\frac{i}{e^2}\bar{\lambda}_{-}\del_{+}\sigma
-T\bar{\lambda}_{+}\non
\so \qquad 
-\frac{k}{2}\chi_{-}(D_0+D_1)\bar{p}+i\frac{k}{2}\chi_{+}\sigma 
+i\frac{k}{2}\bar{\chi}_{+}F_P\Bigr]\non
\so \quad +
\ep_{-}\Bigl[
-\psi_{+}(D_0-D_1)\bar{\phi}+i\bar{\sigma}\bar{\phi}\psi_{-}
-i\bar{\psi}_{-}F-\frac{i}{e^2}\bar{\lambda}_{+}\del_{-}\bar{\sigma}
+\bar{T}\bar{\lambda}_{-}\non
\so \qquad 
-\frac{k}{2}\chi_{+}(D_0-D_1)\bar{p}+i\frac{k}{2}\chi_{-}\bar{\sigma}
-i\frac{k}{2}\bar{\chi}_{-}F_P\Bigr]\non
\so \quad +
\bep_{+}\Bigl[
\bar{\psi}_{-}(D_0+D_1){\phi}+i\bar{\sigma}{\phi}\bar{\psi}_{+}
+i{\psi}_{+}\bar{F}+\frac{i}{e^2}{\lambda}_{-}\del_{+}\bar{\sigma}
+\bar{T}{\lambda}_{+}\non
\so \qquad 
+\frac{k}{2}\bar{\chi}_{-}(D_0+D_1){p}+i\frac{k}{2}\bar{\chi}_{+}
\bar{\sigma}
+i\frac{k}{2}{\chi}_{+}\bar{F}_P\Bigr]\non
\so \quad +
\bep_{-}\Bigl[
\bar{\psi}_{+}(D_0-D_1){\phi}+i{\sigma}{\phi}\bar{\psi}_{-}
-i{\psi}_{-}\bar{F}-\frac{i}{e^2}{\lambda}_{+}\del_{-}{\sigma}
-{T}{\lambda}_{-}\non
\so \qquad 
+\frac{k}{2}\bar{\chi}_{+}(D_0-D_1){p}+i\frac{k}{2}\bar{\chi}_{-}
{\sigma}
-i\frac{k}{2}{\chi}_{-}\bar{F}_P\Bigr]\,,\label{boundary} 
\end{eqnarray}
where we used
\begin{eqnarray}
\so \qquad T=|\phi |^2+\frac{k}{2}(p+\bar{p})
+\frac{1}{2e^2}(D+iv_{01})-t\,.\nom
\ea

There are two types of boundary conditions 
for four susy parameters $(\ep_{+},\ep_{-},\bep_{+},\bep_{-})$.
The one is the A-type boundary, which is realized by setting $\ep_{+}=\bep_{-}=\ep$ 
and $\bep_{+}=\ep_{-}=\bep$ on the boundary $x^1=0$
\ba
\so \theta^{+}=-\bar{\theta}^{-}=\theta\,,\,\,\bar{\theta}^{+}=
-{\theta}^{-}=\bar{\theta}\,,\non
\so \bar{\Phi}e^V\Phi =c_1\,,\,\,
 P+\bar{P}+V=c_2\,,\,\,\Sigma =0\, \quad (c_1, c_2:\mbox{constant}).\nom
\ea
These lead to Dirichlet boundary conditions 
for $|\phi|$, $\mbox{Re}(p)$ and $\sigma$
\ba
\so |\phi |^2=c_1\,,\,\,\psi_{+}\bar{\phi}+\bar{\psi}_-\phi =0\,,\,\,
\psi_{-}\bar{\phi}+\bar{\psi}_+\phi =0\,,\non
\so \bar{\phi}(F+iD_1\phi)+\phi (\bar{F}-iD_1\bar{\phi})
+\bar{\psi}_{+}\psi_{+}-\bar{\psi}_{-}\psi_{-}=0\,,\non
\so p+\bar{p}=c_2\,,\,\,
\chi_{+}+\bar{\chi}_{-}=0\,,\,\,
\bar{\chi}_{+}+\chi_{-}=0\,,\,\,
F_P+\bar{F}_P+i(D_1p-D_1\bar{p})=0\,,\non
\so \sigma =0\,\,,\,\,
 \lambda_{+}+\bar{\lambda}_{-}=0\,,\,\,
\bar{\lambda}_{+}+\lambda_{-}=0\,.\,\,\nom
\ea
The other is the B-type boundary, which is realized by setting 
$\ep_{+}=-\ep_{-}=\ep $ and $\bep_{+}=-\bep_{-}=\bep$.
For $\Phi$ and $\Sigma$, 
the boundary conditions are given by
\ba
\so \theta^{+}=\theta^{-}=\theta\,,\,\,\bar{\theta}^{+}=
\bar{\theta}^{-}=\bar{\theta}\,,\non
\so {\cal D}_{\pm}=e^{-V}D_{\pm}e^{V}\,,\quad
{\cal D}_{+}\Phi ={\cal D}_{-}\Phi\,,\,\,\quad
\Sigma =\bar{\Sigma}\,,\nom
\ea
which are gauge invariant formulae and 
expressed in the component fields as
\ba
\so \sigma =\bar{\sigma}\,,\,\,\lambda_{+}+\lambda_{-}=0\,,\,\,
\bar{\lambda}_{+}+\bar{\lambda}_{-}=0\,\,,\,\,
-2v_{01}=\del_1(\sigma +\bar{\sigma})\,,\non
\so \psi_{+}=\psi_{-}\,,\,\,F=0\,,\,\,
D_1\phi =0\,,\,\,D_1(\psi_{+}+\psi_{-})=0\,.\label{bdy1}
\ea
The condition for 
$\mbox{Im}(\sigma)$
is a Dirichlet type,
and the conditions for 
$\mbox{Re}(\sigma)$ and $\phi$
have modified forms of Neumann types in the presence of 
vector field $v_a$ $(a=0,1)$.
The B-type boundary condition
for the remaining superfield $P$ is 
realized by imposing $D_{+}(P+V)=D_{-}(P+V)$.
This equation is gauge invariant 
since $P$, $V$ transform as Eqs.(\ref{gauge}), 
and $\bar{\Lambda}$ is the
anti-chiral superfield satisfying $D_{\pm}\bar{\Lambda}=0$.
This boundary condition can be written down in the component fields
\ba
\so \chi_{+}=\chi_{-}\,,\,\,F_P=0\,,\,\,
D_1p=0\,\,,\del_1(\chi_{+}+\chi_{-})=0\,.\label{bdy2}
\ea

In the case of B-type,
we can construct the following boundary interaction terms for $\Sigma =\{(x^0,x^1)\,;\,x^1 \geq 0\}$
to cancel the boundary terms in Eq.(\ref{boundary}) after susy transformation
\ba
\so S_{bdy}
=\frac{1}{4\pi}
\int_{\del\Sigma}dx^0\,
\Bigl[
i(\ps_{+}\bps_{-}+\bps_{+}\ps_{-})
+i\frac{k}{2}(\chi_{+}\bar{\chi}_{-}-\chi_{-}\bar{\chi}_{+})\non
\so \quad
+i(\sigma -\bar{\sigma})\left(|\phi
|^2+\frac{k}{2}(p+\bar{p})\right) 
-\frac{1}{2e^2}
\left\{
\del_1|\sigma |^2+2\mbox{Im}(\sigma (D+iv_{01}))
\right\} -i(t\sigma -\bar{t}\bar{\sigma})\Bigr]\,.\nom
\ea
When we plug the boundary conditions (\ref{bdy1}),(\ref{bdy2}) into 
this boundary action, $S_{bdy}$ vanishes for $\theta =0$. 
However 
interaction terms on the boundary are induced for non-zero $\theta$
\ba
\so S_{bdy}=\frac{-\theta}{4\pi}\int_{\del\Sigma}dx^0\,
(\sigma +\bar{\sigma})
=\frac{1}{4\pi}\theta \int_{\del\Sigma}dx^0\,
\left(|\phi |^2+\frac{k}{2}\right)^{-1}
(\bar{\psi}_{+}\psi_{-}+\bar{\psi}_{-}\psi_{+})\,,\label{sig}
\ea
which corresponds to Eq.(\ref{boundary_int}) in the nonlinear sigma model. 


We can easily generalize the above model to the one with 
$N$ chiral superfields $\Phi_i$ $(i=1,2,\cdots ,N)$, 
$M$ vector superfields $V_{\ell}$ ($\ell =1,2,\cdots ,M$) and 
chiral superfields $P_{\ell}$  ($\ell =1,2,\cdots ,M$).
The Lagrangian is defined by
\begin{eqnarray}
&&L=L_K+L_g+L+P+L_{\Sigma}\,,\non
\so L_K=\sum_{i}\Biggl[
D_0\bar{\phi}_iD_0\phi_i -D_1\bar{\phi}_iD_1\phi_i 
+\frac{i}{2}\bar{\psi}_{-i}(D_0+D_1){\psi}_{-i}
+\frac{i}{2}{\psi}_{-i}(D_0+D_1)\bar{\psi}_{-i}\non
\so \qquad 
+\frac{i}{2}\bar{\psi}_{+i}(D_0-D_1){\psi}_{+i}
+\frac{i}{2}{\psi}_{+i}(D_0-D_1)\bar{\psi}_{+i}
+\sum_{\el}Q^{\el}_iD_{\el}|\phi_i|^2
+|F_i|^2\non
\so \qquad -\sum_{\el}(Q^{\el}_i)^2|\sigma_{\el} |^2|\phi_i |^2
-\sum_{\el} Q^{\el}_i\sigma_{\el} \bar{\psi}_{-i}\psi_{+i}
-\sum_{\el}Q^{\el}_i\bar{\sigma}_{\el}\bar{\psi}_{+i}\psi_{-i}\non
\so \qquad 
-i\sum_{\el}Q^{\el}_i\bar{\phi}_i(\lambda_{-\el}\psi_{+i}
-\lambda_{+\el}\psi_{-i})
-i\sum_{\el}Q^{\el}_i
{\phi}_i(\bar{\psi}_{-i}\bar{\lambda}_{+\el}
-\bar{\psi}_{+i}\bar{\lambda}_{-\el})
\Biggr]
\,,\non
\so L_g=\sum_{\el}\frac{1}{2e_{\el}^2}
\Bigl(
\del_0\bar{\sigma}_{\el}\del_0\sigma_{\el}
-\del_1\bar{\sigma}_{\el}\del_1\sigma_{\el}
+i\lambda_{+\el}\del_{-}\bar{\lambda}_{+\el}
+i\bar{\lambda}_{+\el}\del_{-}{\lambda}_{+\el}
+i\lambda_{-\el}\del_{+}\bar{\lambda}_{-\el}\non
\so \qquad 
+i\bar{\lambda}_{-\el}\del_{+}{\lambda}_{-\el}
+D^2_{\el}+v_{01,\el}^2 \Bigr)\,,\non
\so L_P=\sum_{\el}
\frac{k_{\el}}{2}
\Bigl(
D_0\bar{p}_{\el}D_0p_{\el}-D_1\bar{p}_{\el}D_1p_{\el}
+i\chi_{+\el}\del_{-}\bar{\chi}_{+\el}
+i\bar{\chi}_{+\el}\del_{-}{\chi}_{+\el}\non
\so \qquad 
+i\chi_{-\el}\del_{+}\bar{\chi}_{-\el}
+i\bar{\chi}_{-\el}\del_{+}{\chi}_{-\el}
+i\chi_{+\el}\lambda_{-\el}-i\chi_{-\el}\lambda_{+\el}\non
\so \qquad 
+i\bar{\chi}_{+\el}\bar{\lambda}_{-\el}
-i\bar{\chi}_{-\el}\bar{\lambda}_{+\el}
+|F_{P,\el}|^2-|\sigma_{\el} |^2+D_{\el}(p_{\el}+\bar{p}_{\el})
\Bigr)\,,\non
\so L_{\Sigma}=
\sum_{\el}(-r_{\el}D_{\el}+\theta_{\el} 
v_{01,\el})\,, \nn
\end{eqnarray}
where we set
\ba
\so D_a\phi_i =(\del_a+iQ^{\el}_{i}v_{a,\el})\phi_i\,,\,\,
D_a\bar{\phi}_{\el}=
(\del_a-iQ^{\el}_iv_{a,\el})\bar{\phi}_i\,,\non
\so D_ap_{\el}=\del_ap_{\el}+iv_{a,\el}\,,\,\,
D_a\bar{p}_{\el}=\del_a\bar{p}_{\el}-iv_{a,\el}\,,\,\,(a=0,1)\, \nn\\
&&t_{\el}=r_{\el}-i\theta_{\el}\,.\nn
\ea
The associated boundary conditions are 
summarized:
\begin{itemize}
\item B-type boundary $\ep_{+}=-\ep_{-}$\,;
\ba
\so \theta^{+}=\theta^{-}\,,\,\,\bar{\theta}^{+}=
\bar{\theta}^{-}\,,\non
\so \Sigma_{\el}=\bar{\Sigma}_{\el}\,, \quad
D_{+}(P_{\el}+V_{\el})=D_{-}(P_{\el}+V_{\el})\,,\quad
{\cal D}_{+}\Phi_i={\cal D}_{-}\Phi_i\,,\nom
\ea
\item A-type boundary $\ep_{+}=+\bep_{-}$\,;
\ba
\so \theta^{+}=-\bar{\theta}^{-}\,,\,\,
\bar{\theta}^{+}=-{\theta}^{-}\,,\,\,\non
\so P_{\el}+\bar{P}_{\el}+V_{\el}=\hat{c}_{\ell}\,,\quad
 \bar{\Phi}_ie^{Q_i\cdot V}\Phi_i=c_{i}\,,\quad
\Sigma_{\ell}=0\,.\nom
\ea
\end{itemize}
By eliminating the auxiliary fields in the  potential terms, several
 constraints are obtained;
\ba
 -\frac{D_{\el}}{e_{\el}^2}
=\sum_{i}Q^{\ell}_i
|\phi_i |^2+\frac{k_{\el}}{2}(p_{\el}+\bar{p}_{\el})-r_{\el}\,,\quad
F_i=F_{P,\el}=0\,.\nom
\ea
In the limit $e^2 \rightarrow \infty$, 
the model is reduced to a nonlinear sigma model and $\sigma_{\ell}$'s are written down 
\ba
\so \sigma_{\el} =-\left(\sum_i (Q^{\el}_i)^2|\phi_i |^2
+\frac{k_{\el}}{2}\right)^{-1}\sum_i Q^{\el}_i
\bar{\psi}_{+i}\psi_{-i}\,,\nn\\
&& \bar{\sigma}_{\el}=-\left(\sum_i (Q^{\el}_i)^2
|\phi_i |^2+\frac{k_{\el}}{2}\right)^{-1}\sum_i Q^{\el}_i
\bar{\psi}_{-i}\psi_{+i}\,,\nom
\ea
which leads to the boundary terms as in Eq.(\ref{sig}).
\ba
\so \sigma_{\el} +\bar{\sigma}_{\el}=
-\left(\sum_i (Q^{\el}_i)^2|\phi_i |^2+\frac{k_{\el}}{2}\right)^{-1}
\sum_i Q^{\el}_i
(\bar{\psi}_{+i}\psi_{-i}+\bar{\psi}_{-i}\psi_{+i})\,.\nom
\ea

Here we investigate the FI-term.
For simplicity, let us consider the case of $\ell =1$.
In the $e^2 \rightarrow\infty $ limit, 
the explicit forms of the gauge field $v_0$, $v_1$ is obtained,
\ba
\so v_{0}=
\frac{1}{2}\left(\sum_iQ_i^2|\phi_i|^2
+\frac{k}{2}\right)^{-1}
\Bigl[i\sum_iQ_i(\bar{\phi}_i\del_0\phi_i-\phi_i\del_0\bar{\phi}_i)
+i\frac{k}{2}\del_0(p-\bar{p})\non
\so \qquad +\sum_iQ_i(\bps_{+i}\ps_{+i}+\bps_{-i}\ps_{-i})
\Bigr]\,,\non
\so v_{1}=
\frac{1}{2}\left(\sum_iQ_i^2 |\phi_i|^2
+\frac{k}{2}\right)^{-1}
\Bigl[i\sum_i Q_i(\bar{\phi}_i\del_1\phi_i-\phi_i\del_1\bar{\phi}_i)
+i\frac{k}{2}\del_1(p-\bar{p})\non
\so \qquad +\sum_i Q_i(\bps_{+i}\ps_{+i}-\bps_{-i}\ps_{-i})
\Bigr]\,.\nom
\ea
The configuration represented by the above $v_0$, $v_1$ contributes to the FI-terms, in particular to the theta-term
\ba
\so S_{\Sigma}=\frac{1}{2\pi}\int dx^0dx^1\,(-rD
+\theta v_{01})\non
\so \qquad =
\frac{\theta}{2\pi}\int dx^0dx^1\,
\left(\sum_i Q_i^2|\phi_i|^2+\frac{k}{2}\right)^{-1}
\sum_i \Bigl[
iQ_i\{(\hat{D}_1\phi_i)(\hat{D}_0\bar{\phi}_i)
-(\hat{D}_1\bar{\phi}_i)(\hat{D}_0{\phi}_i)\}\Bigr]\non
\so \qquad 
+\frac{\theta}{2\pi}\int dx^0dx^1\,\sum_i
\Bigl[
-Q_i\del_{+}
\left\{\left(\sum_j Q_j^2|\phi_j|^2+\frac{k}{2}\right)^{-1}
\psi_{-i}\bps_{-i}\right\}\non
\so \qquad 
+Q_i\del_{-}
\left\{\left(\sum_j Q_j^2|\phi_j|^2+\frac{k}{2}\right)^{-1}
\ps_{+i}\bps_{+i}\right\}
\Bigr]+\cdots \,,\non
\so 
\hat{D}_a\phi_i=\del_a\phi_i+iQ_i\hat{v}_a\phi_i\,,\non
\so \hat{v}_a=
\frac{1}{2}\left(\sum_iQ_i^2|\phi_i|^2
+\frac{k}{2}\right)^{-1}
\Bigl[iQ_i
(\bar{\phi}_i\del_a\phi_i-\phi_i\del_a\bar{\phi}_i)
+i\frac{k}{2}\del_a(p-\bar{p})
\Bigr]\,.\nom
\ea
Note that the elimination of gauge potential induces 
anti-symmetric fields.

Now let us introduce the linear sigma model realizing  
the K\"ahler geometry in $2n$-dimensional target space \cite{Hori:2001ax} 
discussed in section \ref{section_generalize}.
The action contains
$n$ chiral superfields $\Phi$, one chiral superfield $P$ and 
one U($1$) gauge field $V$
\ba
&&L=L_K+L_P+L_g\,,\non
&& L_K=\sum_{i=1}^n\int d^4\theta\,\bar{\Phi}_ie^V\Phi_i\,,\quad
L_P=\frac{k}{4}\int d^4\theta\,(P+\bar{P}+V)^2\,,\quad
L_g=-\frac{1}{2e^2}\int d^4\theta\,\bar{\Sigma}\Sigma\,.\nom
\ea
There are $n$ complex scalars $\phi_i$'s and
one complex scalar $p$, 
but there is a D-flatness condition
\ba
\so - \frac{D}{e^2}=\sum_{i=1}^n|\phi_i|^2+\frac{k}{2}(p+\bar{p})=0\,,\nom
\ea
and the imaginary part of $p$ can be gauged away. 
So the remaining degrees of freedom is 
$2n=2n+2-1-1$ and 
dimension of the target space geometry is $2n$.
The target space of the sigma model is 
characterized by its kinetic terms and 
one can get the metric of the spacetime:
\ba
\so ds^2=2\sum_{i=1}^n d\phi_id\bar{\phi}_i+kdpd\bar{p}\,.\nom
\ea
We parameterize $\phi_i$'s as
\ba
\so \sum_{i=1}^n|\phi_i|^2=r^2\,,\,\,\,\,\label{wa}\quad
 \phi_i= r F^{-1/2}e^{\frac{i}{n}\varphi}
\cdot w_i\quad (i=1,2,\cdots n\,;\,w_n=1)\,,\quad
 F=1+\sum_{i=1}^{n-1}|w_i|^2\,.\nom
\ea
{}From the D-flatness condition, one has to impose 
a relation  $r^2+k\mbox{Re}(p)=0$ for the real part of $p$. 
On the other hand, the imaginary part $\mbox{Im}(p)$ 
is related to $\varphi$ through gauge transformation;
$\varphi \rightarrow \varphi +n\alpha$, 
$\mbox{Im}(p)\rightarrow \mbox{Im}(p)+\alpha$. 
From these relations one can write the spacetime metric\cite{Hori:2001ax}
\ba
&&ds^2=2\hat{f}(r)dr^2+\frac{2r^2}{n^2\hat{f}(r)}(d\varphi -nA)^2
+2r^2ds_{FS}^2\,,\non
\so \hat{f}(r)=1+\frac{2r^2}{k}\, \quad
A=-\frac{i}{2}F^{-1}\sum_{i=1}^{n-1}
(w_id\bar{w}_j-\bar{w}_idw_j)\,,\non
\so ds_{FS}^2=F^{-1}\sum_{i,j}^{n-1}(\delta_{ij}-
F^{-1}w_i\bar{w}_j)d\bar{w}_idw_j\,.
\ea
It is a K\"ahler metric and one can consider $N=(2,2)$ nonlinear sigma models
with this target space.
By the same procedure as the analysis of the nonlinear sigma model 
in section \ref{section_generalize},
we can get A-, B-type susy boundary conditions.

Here we consider the linear sigma model.
For the A-type case, one gets boundary conditions
$|\phi_i|^2=c_i \geq 0\,(i=1,2,\cdots ,n)$, 
which represents an $n$-dimensional object in the geometry.
The $r$ is determined uniquely by 
$\sum_{i=1}^{n}c_i=r^2$. 
Thus the boundary condition for the $r$ direction is 
a Dirichlet type at least classically.
This boundary conditions also
impose the 
constraints $F^{-1}|w_i|^2=c_ir^{-2}$
for the absolute values of $w_i$ ($i=1,2,\cdots,n$) of the base space CP${}^{n-1}$
on each local coordinate patch.
But we can choose phases (angular coordinates)
of $w_i$'s arbitrarily and 
the object is extended in the angular directions.
For the 2-dim black hole, this configuration 
corresponds to Fig. \ref{figure:axial} (i).
This consideration might be true only at classical level (discussion of the 
sigma model at tree level).
The geometry discussed now is classical and changes into another one 
through RG flow. 
The resulting metric is thought to the one
studied in section \ref{section_generalize}.
Comparing these two metrics, we see that 
both the metrics have similar forms. 
The radial part $\hat{f}(r)$ is
changed into $g_n(Y)$,
while the $w_i$'s are coordinates of the base CP${}^{n-1}$ 
space in both metrics.

Also there exists an effect of a dilaton and 
the brane tension restricts the shape of the 
susy objects. 
The tension of the brane is proportional to 
$e^{nY/k}$ and in particular  the A-type branes considered above 
should shrink to $r=0$.  
The radial parts $Y$, $|w^i|^2$'s and 
angular parts $\theta$, $\mbox{arg}\,w_i$'s 
play a complementary role in the K\"ahler form 
because this geometry has the U($n$) isometry. 
So it is natural to consider non-compact objects 
extended to the radial parts.
In the case of $2$-dim black hole
there are the D-branes extended 
in the noncompact direction (see Fig. \ref{figure:axial} (ii) etc). 
Therefore noncompact branes corresponding to them should exist in this model.

 
\section{Conclusions}

In this paper we have studied  the D-branes of the $N=2$ supersymmetric sigma model. 
We made it clear that 
the nonlinear sigma model has D0,D1,D2-branes in the $2$-dim target space  
by considering A-,B-type supersymmetric boundary conditions on the worldsheet,
and achieved the boundary interaction term added to the action. 
Especially in the case of black hole metric, 
we obtained the geometrical descriptions for the D-branes consistent with the result 
in the analysis of conjugacy classes.

Furthermore we investigated 
the linear sigma model realizing dynamics on the generalized metrics with U$(n)$ isometry.
We considered two types of boundary conditions and  
obtained the D-branes consistent with those in the $2$-dim case. 

\section*{Acknowledgments}

The work of K.S. is 
supported in part by the Grant-in-Aid from the Ministry of 
Education, Science, Sports and Culture of Japan ($\sharp$ 14740115).

\appendix

\section{Nonlinear Sigma Model with F-term}

In section \ref{section_nonlinear}
we investigated nonlinear sigma models without F-term.
It is possible to have similar discussion
even in the presence of F-term (superpotential $W(\phi)$). 

The action is 
\beq
S &=& \frac{k}{4 \pi} \int_{\Sigma} d^2 z \Bigl[
\del\bar{\del}K \cdot(\del_{+}\phi\del_{-}\phib+
\del_{-}\phi\del_{+}\phib)\non
&& -i\chi^{+}\chi^{-}\left(\del_{+}\phi\cdot \frac{\del^2 \bar{\del}K}{\del\bar{\del}K}
-\del_{+}\phib \cdot \frac{\del \bar{\del}^2 K}{\del\bar{\del}K}\right)
-i\lambda^{+}\lambda^{-}\left(\del_{-}\phi\cdot \frac{\del^2 \bar{\del}K}{\del\bar{\del}K}
-\del_{-}\phib \cdot \frac{\del \bar{\del}^2K}{\del\bar{\del}K}\right)\non
&& +2\lambda^{+}\lambda^{-}\chi^{+}\chi^{-}
\frac{\del^2 \bar{\del}^2 K}{(\del\bar{\del}K)^2}  +i (\lambda^{+}\del_{-}\lambda^{-}
+\lambda^{-}\del_{-}\lambda^{+}
+\chi^{+}\del_{+}\chi^{-}
+\chi^{-}\del_{+}\chi^{+})\non
&&+\frac{1}{2}\ \del\bar{\del}K \cdot F \bar{F}-\bar{F} \frac{\del^2\bar{\del}K}{\del\bar{\del}K}  \cdot \lambda^{+}\chi^{+}
+F \frac{\del\bar{\del}^2K}{\del\bar{\del}K}  \cdot \lambda^{-}\chi^{-} \nn\\
&& -(\del\bar{\del}K)^{-1}\del^2W\cdot \lambda^{+}\chi^{+}
+(\del\bar{\del}K)^{-1}\bar{\del}^2\bar{W}\cdot \lambda^{-}\chi^{-}
+\frac{1}{2}F\del W+\frac{1}{2}\bar{F}\bar{\del}\bar{W}
\Bigr]\,.\nom
\eeq
If $\del\bar{\del}K$ does not blow up, 
this is reduced to  a canonical form of $N=(2,2)$ supersymmetric nonlinear sigma model
by the redefinition of fermionic fields 
(\ref{def_fermion}).
The supersymmetry transformation is expressed as
\beq
&&\delta \phi =\sqrt{2} (\del \bar{\del} K)^{-1/2} (\epsilon_{+}\chi^{+}-\epsilon_{-}\lambda^{+}),  \quad 
\delta \phib =\sqrt{2}(\del \bar{\del}K)^{-1/2}(-\bar{\epsilon}_{+}\chi^{-}+\bar{\epsilon}_{-}\lambda^{-})\,,\non
&&\delta F= -2 
\sqrt{2} i \bar{\epsilon}_{+} \del_{-} \left( ( \del \bar{\del}K)^{-1/2}\lambda^{+} \right)
-2 \sqrt{2}i\bar{\epsilon}_{-}\del_{+} \left( (\del \bar{\del}K)^{-1/2}\chi^{+} \right)\,,\non
&&\delta \bar{F}=-2 \sqrt{2}i{\epsilon}_{+}\del_{-} \left( (\del \bar{\del}K)^{-1/2}\lambda^{-} \right)
-2\sqrt{2}i{\epsilon}_{-}\del_{+} \left( (\del \bar{\del}K)^{-1/2}\chi^{-} \right)\,,\non
&&\delta \lambda^{+}=\sqrt{2}(\del \bar{\del}K)^{1/2}\bar{\epsilon}_{-}
\left(i\del_{+}\phi -\frac{\del \bar{\del}^2 K}{2(\del \bar{\del}K)^{2}}\cdot 
\lambda^{+}\lambda^{-}\right)
\non
&&\qquad +\frac{1}{\sqrt{2}}(\del \bar{\del}K)^{1/2}
\Biggl[
\epsilon_{+}\left(F+\frac{\del^2 \bar{\del}K}{(\del \bar{\del}K)^{2}}\cdot \chi^{+}\lambda^{+}\right)
-\frac{\del \bar{\del}^2 K}{(\del \bar{\del}K)^{2}}\cdot \bar{\epsilon}_{+}\chi^{-}\lambda^{+}
\Biggr], \nn \\
&&\delta \chi^{+}=\sqrt{2}(\del \bar{\del}K)^{1/2}\bar{\epsilon}_{+}
\left(-i\del_{-}\phi
+\frac{\del \bar{\del}^2 K}{2(\del \bar{\del}K)^{2}}\cdot \chi^{+}\chi^{-}\right)
\non
&&\qquad +\frac{1}{\sqrt{2}}(\del \bar{\del}K)^{1/2}
\Biggl[
\epsilon_{-}\left(F+\frac{\del^2 \bar{\del}K}{(\del \bar{\del}K)^{2}} \cdot \chi^{+}\lambda^{+}\right)
-\frac{\del \bar{\del}^2 K}{(\del \bar{\del}K)^{2}}
\cdot \bar{\epsilon}_{-}\chi^{+}\lambda^{-}
\Biggr], \nn \\
&&\delta \lambda^{-}=\sqrt{2}(\del \bar{\del}K)^{1/2}\epsilon_{-}
\left(-i\del_{+}\phib
-\frac{\del^2 \bar{\del} K}{2(\del \bar{\del}K)^{2}}\cdot \lam^{+}\lam^{-}\right)
\non
\non
&&\qquad +\frac{1}{\sqrt{2}}(\del \bar{\del}K)^{1/2}
\Biggl[
\bar{\epsilon}_{+}\left(\bar{F}-\frac{\del \bar{\del}^2 K}{(\del \bar{\del}K)^{2}} \cdot \chi^{-}\lambda^{-}\right)
+\frac{\del^2 \bar{\del} K}{(\del \bar{\del}K)^{2}}
\cdot {\epsilon}_{+}\chi^{+}\lambda^{-}
\Biggr],\nn\\
&&\delta \chi^{-}=\sqrt{2}(\del \bar{\del}K)^{1/2}\epsilon_{+}
\left(i\del_{-}\phib
+\frac{\del^2 \bar{\del} K}{2(\del \bar{\del}K)^{2}}\cdot \chi^{+}\chi^{-}\right)
\non
&&\qquad +\frac{1}{\sqrt{2}}(\del \bar{\del}K)^{1/2}
\Biggl[
\bar{\epsilon}_{-}\left(\bar{F} -\frac{\del \bar{\del}^2 K}{(\del \bar{\del}K)^{2}} \cdot \chi^{-} \lam^-\right)
+\frac{\del^2 \bar{\del} K}{(\del \bar{\del}K)^{2}}
\cdot {\epsilon}_{-}\chi^{-} \lam^+\Biggr]\,,\nom
\eeq
where $\ep_{+}$,$\ep_{-}$,$\bep_{+}$,$\bep_{-}$ are 
susy parameters.
Under the above transformation, we obtain 
the variation of the action
\beq
\delta S &=& \frac{k}{4 \pi} \frac{1}{2}
\int_{\del \Sigma} d\xi_1 \Bigl[ 
-i \sqrt{2}(\del \delb K)^{1/2}
\left( 
\epsm \lamp \delm \phib- \epsbm \lamm \delm \phi+ \epsp \chip \delp \phib - \epsbp \chim \delp \phi
\right) \nn \\
&&+\frac{1}{\sqrt{2}}(\del \delb K)^{-1/2}
\left(\epsm \chim \delb \bar{W} + \epsbm \chip \del W 
-\epsp \lamm \delb \bar{W} - \epsbp \lamp \del W 
\right)\nn\\
&&
+\frac{1}{\sqrt{2}}(\del\bar{\del}K)^{-1/2}
(\ep_{-}\chi^{-}-\ep_{+}\lambda^{-})
\left(F\cdot \del\bar{\del}K-2\frac{\del^2\bar{\del}K}{\del\bar{\del}K}
\lambda^{+}\chi^{+}+\bar{\del}\bar{W}\right)\nn\\
&&
+\frac{1}{\sqrt{2}}(\del\bar{\del}K)^{-1/2}
(\bep_{-}\chi^{+}-\bep_{+}\lambda^{+})
\left(\bar{F}\cdot \del\bar{\del}K-2\frac{\del\bar{\del}^2K}{\del\bar{\del}K}
\chi^{-}\lambda^{-}+\del W\right)
\Bigr].\nn
\eeq
When one uses equations of motion of $F$ and $\bar{F}$, 
the last two lines vanish. 
The second line  
contains contributions of $W$ and $\bar{W}$. 
But these terms vanish 
at the  singular point given by $(\del\bar{\del}K)^{-1}=0$.

We shall write down susy boundary conditions;
\begin{enumerate}
\item A-type \, $\ep_{+}=\bep_{-}$, $\bep_{+}=\ep_{-}$\,;
\ba
&&\lamm \pm \chip = \lamp \pm \chim =0, \nn \\
&&\delm \phi \pm \delp \phib =0, \quad \delm \phib \pm \delp \phi =0, \quad
\del W \pm \delb \bar{W}=0,\nn
\ea
\item B-type \, $\ep_{+}=-\ep_{-}$, $\bep_{+}=-\bep_{-}$\,;
\subitem 
\ba
&&\lambda^{+}-\chip=\lambda^{-}-\chim=0\,,\,\,\non
&&\del_2\phi =\del_2\bar{\phi}=0\,,\quad
\del W=\bar{\del}{\bar{W}}=0\,,\nom
\ea
or
\subitem
\ba
&&\lambda^{+}+\chip=\lambda^{-}+\chim=0\,,\,\,\non
&&\del_1\phi =\del_1\bar{\phi}=0\,.\,\,\nom
\ea
\end{enumerate}

\end{document}